\documentclass{article}
\usepackage{dcase2020,amsmath,graphicx,url,times,booktabs}
\usepackage{amssymb}
\usepackage{enumerate}
\usepackage{graphicx}
\usepackage{multirow}
\usepackage{arydshln}
\usepackage{color,xcolor}
\usepackage{float}
\usepackage{bm}
\usepackage{comment}
\usepackage{cite}
\usepackage{url}
\usepackage{amsmath}
\usepackage{algorithmic}
\usepackage{array}
\usepackage{colortbl}
\usepackage{here}

\newcommand{\bi}[1]{\ensuremath{\boldsymbol{#1}}}   %  Definition of Bold Itaric
\newlength\savedwidth
\newcommand{\wcline}[1]{\noalign{\global\savedwidth\arrayrulewidth\global\arrayrulewidth 1.0pt} \cline{#1}
\noalign{\global\arrayrulewidth\savedwidth}}

\title{Sound Event Detection Using Duration Robust Loss Function}

\name{Daichi Akiyama$^{\dagger \ast}$, Keisuke Imoto$^{\dagger \diamondsuit \ast}$, Noriyuki Tonami$^{\dagger}$, Yuki Okamoto$^{\dagger}$,}
\secondlinename{Ryosuke Yamanishi$^{\dagger \ddagger}$, Takahiro Fukumori$^{\dagger}$, Yoichi Yamashita$^{\dagger}$\thanks{$^{\ast}$ These authors contributed equally to this work.}
}
\address{$^{\dagger}$\hspace{1pt}Ritsumeikan University, $^{\diamondsuit}$\hspace{1pt}Doshisha University, $^{\ddagger}$\hspace{1pt}Kansai University}
\begin{document}
\ninept
\maketitle
\begin{sloppy}
%---------------------------------------------------
%\vspace{0pt}
\begin{abstract}
%\vspace{0pt}
%---------------------------------------------------
%
Many methods of sound event detection (SED) based on machine learning regard a segmented time frame as one data sample to model training.
However, the sound durations of sound events vary greatly depending on the sound event class, e.g., the sound event ``fan'' has a long time duration, while the sound event ``mouse clicking'' is instantaneous.
The difference in the time duration between sound event classes thus causes a serious data imbalance problem in SED. 
In this paper, we propose a method for SED using a duration robust loss function, which can focus model training on sound events of short duration.
In the proposed method, we focus on a relationship between the duration of the sound event and the ease/difficulty of model training.
In particular, many sound events of long duration (e.g., sound event ``fan'') are stationary sounds, which have less variation in their acoustic features and their model training is easy.
Meanwhile, some sound events of short duration (e.g., sound event ``object impact'') have more than one audio pattern, such as attack, decay, and release parts.
We thus apply a class-wise reweighting to the binary-cross entropy loss function depending on the ease/difficulty of model training.
Evaluation experiments conducted using TUT Sound Events 2016/2017 and TUT Acoustic Scenes 2016 datasets show that the proposed method respectively improves the detection performance of sound events by 3.15 and 4.37 percentage points in macro- and micro-Fscores compared with a conventional method using the binary-cross entropy loss function.
%
%---------------------------------------------------
%\vspace{0pt}
\end{abstract}
%\vspace{0pt}
%---------------------------------------------------
%
%---------------------------------------------------
%---------------------------------------------------
%
%
%---------------------------------------------------
%\vspace{0pt}
\section{Introduction}
\label{sec:intro}
%---------------------------------------------------
%\vspace{0pt}
%---------------------------------------------------
Sound event detection (SED) is the task of detecting sound event labels and their onset/offset in an audio recording, where a sound event indicates a type of sound such as ``people talking'' and ``bird singing'' \cite{Imoto_AST2018_01}.
SED plays an important role in realizing various applications using artificial intelligence in sounds, such as automatic life-logging, machine monitoring, automatic surveillance, media retrieval, and biomonitoring systems \cite{Imoto_INTERSPEECH2013_01,Koizumi_TASLP2019_01,Koizumi_arXiv2020_01,Ntalampiras_ICASSP2009_01,Jin_INTERSPEECH2012_01,Salamon_PLoSOne2016_01,Okamoto_NCSP2020_01}.

\begin{figure}[t]
%\vspace{0pt}
\centering
\includegraphics[width=0.98\columnwidth]{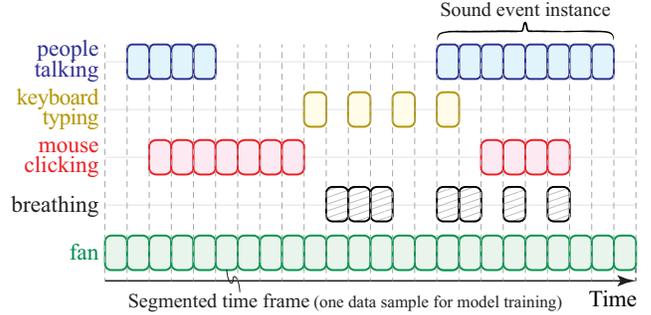}
\vspace{-7pt}
\caption{Examples of durations of sound event instances and number of data samples}
\label{fig:duration}
\vspace{-4pt}
\end{figure}
\begin{table}[t]
%\vspace{0pt}
\small
\caption{Average duration of one sound event instance in datasets used for evaluation experiments (TUT Sound Events 2016, 2017, and TUT Acoustic Scenes 2016 development [12, 13])}
%\cite{Mesaros_EUSIPCO2016_01,Mesaros_DCASE2017_01}
\vspace{-17pt}
\label{tbl:duration}
\begin{center}
\begin{tabular}{lrclr}
\wcline{1-2}\wcline{4-5}
&\\[-9pt]
\multicolumn{1}{c}{\textbf{Sound event}} \!\!&\!\! \textbf{Duration} \!\!&\!\!\!\!&\multicolumn{1}{c}{\!\! \textbf{Sound event}} \!\!&\!\! \textbf{Duration}\\[-1pt]
\wcline{1-2}\wcline{4-5}
&\\[-9pt]
(object) banging \!\!\!\!&\!\!\!\! 0.78 s \!\!&\!\!\!\!&\!\! drawer \!\!\!\!&\!\!\!\! 0.80 s \\
(object) impact \!\!\!\!&\!\!\!\! 0.35 s \!\!&\!\!\!\!&\!\! fan \!\!\!\!&\!\!\!\! 29.99 s \\
(object) rustling \!\!\!\!&\!\!\!\! 2.24 s \!\!&\!\!\!\!&\!\! glass jingling \!\!\!\!&\!\!\!\! 0.80 s \\
(object) snapping \!\!\!\!&\!\!\!\! 0.46 s \!\!&\!\!\!\!&\!\! keyboard typing \!\!\!\!&\!\!\!\! 0.21 s \\
(object) squeaking \!\!\!\!&\!\!\!\! 0.74 s \!\!&\!\!\!\!&\!\! large vehicle \!\!\!\!&\!\!\!\! 14.68 s \\
bird singing \!\!\!\!&\!\!\!\! 7.63 s \!\!&\!\!\!\!&\!\! mouse clicking \!\!\!\!&\!\!\!\! 0.14 s \\
brakes squeaking \!\!\!\!&\!\!\!\! 1.65 s \!\!&\!\!\!\!&\!\! mouse wheeling \!\!\!\!&\!\!\!\! 0.16 s \\
breathing \!\!\!\!&\!\!\!\! 0.43 s \!\!&\!\!\!\!&\!\! people talking \!\!\!\!&\!\!\!\! 4.09 s \\
car \!\!\!\!&\!\!\!\! 6.88 s \!\!&\!\!\!\!&\!\! people walking \!\!\!\!&\!\!\!\! 6.63 s \\
children \!\!\!\!&\!\!\!\! 6.87 s \!\!&\!\!\!\!&\!\! washing dishes \!\!\!\!&\!\!\!\! 4.15 s \\
cupboard \!\!\!\!&\!\!\!\! 0.65 s \!\!&\!\!\!\!&\!\! water tap running \!\!\!\!&\!\!\!\! 5.92 s \\
cutlery \!\!\!\!&\!\!\!\! 0.74 s \!\!&\!\!\!\!&\!\! wind blowing \!\!\!\!&\!\!\!\! 6.09 s \\\wcline{4-5}
dishes \!\!\!\!&\!\!\!\! 1.24 s \!\!&\!\!\!\!&\!\!&\\
\wcline{1-2}
\end{tabular}
\vspace{-5pt}
\end{center}
\end{table}
\begin{figure*}[t!]
%\vspace{0pt}
\centering
\includegraphics[width=1.95\columnwidth]{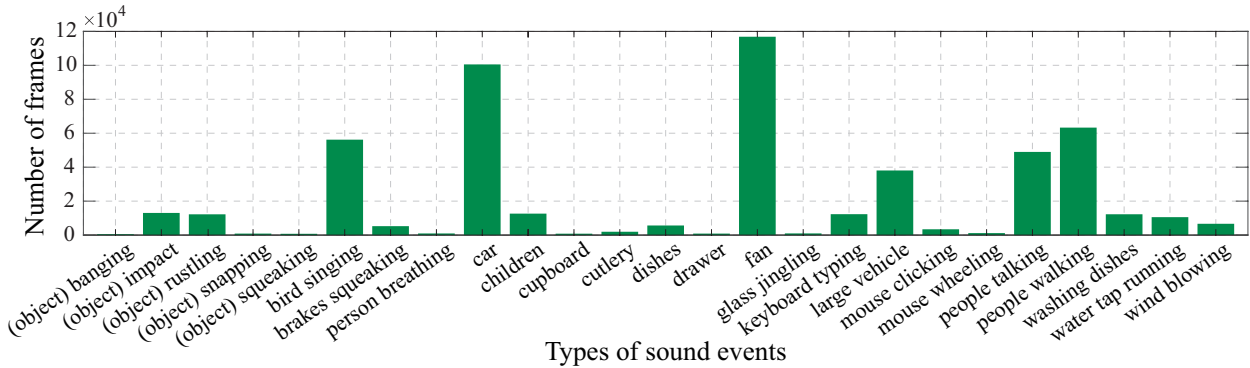}
\vspace{-9pt}
\caption{Numbers of frames of sound events in dataset used for evaluation experiments}
\label{fig:num_frame_01}
%\vspace{0pt}
\end{figure*}

For machine-learning-based SED, many methods using neural networks, such as a convolutional neural network (CNN) \cite{Hershey_ICASSP2017_01}, recurrent neural network (RNN) \cite{Hayashi_TASLP2017_01}, and convolutional recurrent neural network (CRNN) \cite{Cakir_TASLP2017_01}, have been proposed.
In these methods, an audio clip is segmented into short time frames (e.g., 40 ms frame length with 20 ms shift), and each segmented frame is regarded as one data sample when training and evaluating a sound event model.
As shown in Fig.~\ref{fig:duration}, each sound event instance has a different frame length, and the frame length of each instance varies depending on the sound event class.
Table~\ref{tbl:duration} and Fig.~\ref{fig:num_frame_01} respectively show the average duration of one sound event instance and the total frame number of sound events in datasets used for evaluation experiments discussed in Section \ref{sec:experiments} (TUT Sound Events 2016, 2017, and TUT Acoustic Scenes 2016 development \cite{Mesaros_EUSIPCO2016_01,Mesaros_DCASE2017_01}).
In this dataset, the number of frames in sound event ``mouse clicking,'' which has an average length of 0.15 s, is 1,163, while that in sound event ``fan,'' which has an average length of 29.99 s, is 116,837.
Thus, the difference in the time duration between sound events causes a serious data imbalance problem in SED. 
There are some conventional methods of SED for imbalanced data \cite{Chen_INTERSPEECH2019_01,Wang_ICASSP2020_01}.
For instance, Chen and Jin have proposed a method for detecting rare sound events using data augmentation \cite{Chen_INTERSPEECH2019_01}.
Wang \textit{et al.} have proposed a method for few-shot sound event detection based on metric learning \cite{Wang_ICASSP2020_01}.
However, the data imbalance problem caused by the difference in time duration between sound event classes has not been investigated in these works.

In this paper, we address the imbalanced-data SED caused by the difference in time duration using a duration robust loss function.
In a preliminary experiment, it is proven that a simple classwise reweighting of the loss function on the basis of the inverse frequency of sound event occurrences (the details are described in Sec. \ref{Conventional}) is not effective for SED using extremely imbalanced data.
%n
In the proposed method, we instead apply another classwise reweighting approach of the loss function in accordance with on the ease/difficulty of model training.
As shown in Fig.~\ref{fig:spectrogram_01}, this is because many sound events of long duration (e.g., ``fan'' or ``car'') are stationary sounds, which have less variation in their acoustic features and their model training is easy.
We demonstrate that the proposed reweighting approach of the loss function can prevent the sound events of long duration from dominating the model training and improve the performance of SED using a seriously imbalanced dataset.

The rest of this paper is organized as follows.
In Sec. 2, we introduce the conventional SED method, and in Sec. 3, we propose the SED method based on the duration robust loss function.
In Sec. 4, we discuss the SED performance evaluation for an imbalanced training dataset.
In Sec. 5, we conclude this paper.
%
%
%
%---------------------------------------------------
%\vspace{0pt}
\section{Conventional Method}
\label{Conventional}
%\vspace{0pt}
%---------------------------------------------------
Let us consider the training dataset $\mathcal D = \{ ({\bf X}_{1}, {\bf Z}_{1}), ..., ({\bf X}_{l}, {\bf Z}_{l}), ..., $ $({\bf X}_{L}, {\bf Z}_{L}) \hspace{-1pt} \}$.
${\bf X}_{l}$ is an acoustic feature of the $l^{th}$ sound clip and ${\bf Z}_{l} = \{ {\bf z}_{l,1}, ...$ ${\bf z}_{l,n}, ..., {\bf z}_{l,N}\}$ is the sound event label, where ${\bf z}_{l,n} \in \{0, 1\}^{M}$ indicates a multi--hot vector of time frame $n$ in the $l^{th}$ sound clip over the $M$ sound event class.
The goal of SED is to predict sound event labels $\hat{{\bf Z}}$ in an unknown sound using 

\vspace{-8pt}
\begin{align}
\hat{{\bf Z}} = \{ {\bf Z} \in \{0,1\}^{N \times M} | \hspace{1pt} f({\bf X}, {\bi \theta}) \geq \phi \},
\end{align}
%\vspace{0pt}
%

\noindent where $f$, ${\bi \theta}$, and $\phi$ are the model, the model parameter trained using $\mathcal D$, and the detection threshold, respectively.
In the conventional SED, the mel-band energy and mel-frequency cepstral coefficients (MFCCs) are often used as the acoustic features ${\bf X}_{l}$.
As the model $f$, CNN, RNN, or CRNN-based neural network is applied.
The model parameter ${\bi \theta}$ is estimated using the following binary cross-entropy (BCE) loss function $E ({\bi \theta})$ and the backpropagation technique:

\vspace{-10pt}
\begin{align}
{\rm E}_{1} ({\bi \theta}) &= - \! \sum^{N}_{n=1} \! {\Big \{} {\bf z}_{n} \log {\big (} s({\bf y}_{n}) {\big )} \! + \! (1-{\bf z}_{n}) \log {\big (}1-s({\bf y}_{n}){\big )} \! {\Big \}} \nonumber\\[1pt]
&= - \sum^{M}_{m=1} \sum^{N}_{n=1} \! {\Big \{} z_{n,m} \log {\big (} s(y_{n,m}) {\big )} \nonumber\\[0pt]
&\hspace{25pt} + (1 - z_{n,m}) \log {\big (} 1 - s(y_{n,m}) {\big )} {\Big \}},
\label{eq:conv_loss}
%\vspace{0pt}
\end{align}
%\vspace{0pt}
%

\noindent where $s$ and $y_{n,m}$ are the sigmoid function and the output of the network in time frame $n$, respectively.
$z_{n,m}$ is the target label in time frame $n$ and is 1 if acoustic event $m$ is active in time frame $n$ and 0 otherwise.
Note that we omit sound clip index $l$ to simplify the equation.
${\rm E}_{1} ({\bi \theta})$ is actually calculated by summing the binary cross entropy over time frames of all sound clips.
Since the frame length of sound events varies considerably depending on the event class, the model parameter estimation using Eq. (\ref{eq:conv_loss}) leads to the data imbalance problem.
As a result, sound events of long durations overwhelm the model training and those of short durations are likely to be downweighted.

\begin{figure}[t!]
\vspace{5pt}
\hspace{-10pt}
\centering
\includegraphics[width=1.025\columnwidth]{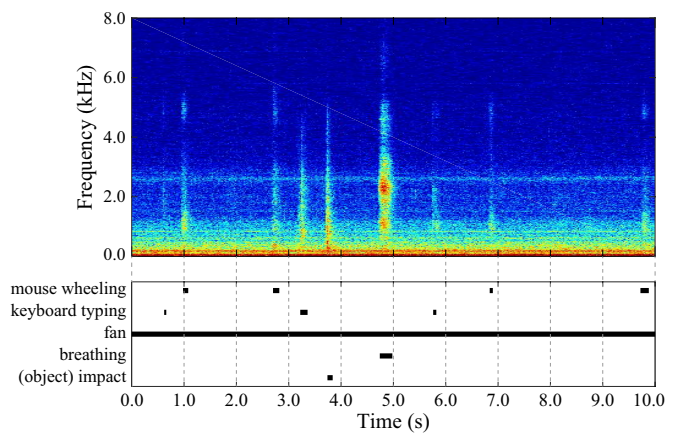}
\vspace{-15pt}
\caption{Spectrogram and sound event label of long/short duration sounds}
\label{fig:spectrogram_01}
\vspace{-20pt}
\end{figure}
\begin{table*}[h]
%\vspace{0pt}
\small
\caption{Sound event detection performance for each event}
\vspace{1pt}
\label{tbl:each_event}
\centering
\scalebox{0.779}[0.779]{
\begin{tabular}{llcccccccccccc}
\wcline{1-14}
&&&&&&&&&&&&&\\[-9pt]
\multicolumn{2}{c}{\multirow{2}{*}{Event}} & (object) & (object) & (object) & (object) & (object) & bird & brakes & \multirow{2}{*}{breathing} & \multirow{2}{*}{car} & \multirow{2}{*}{children} & \multirow{2}{*}{cupboard} & \multirow{2}{*}{cutlery}\\[-1pt]
&& banging & impact & rustling & snapping &  squeaking & singing & squeaking &&&&&\\[0pt]
\wcline{1-14}\\[-9.5pt]
\multirow{2.2}{*}{BCE loss}& \cellcolor[rgb]{0.95,0.95,0.95} Fscore & \cellcolor[rgb]{0.95,0.95,0.95} 0.00\% & \cellcolor[rgb]{0.95,0.95,0.95} 1.03\% & \cellcolor[rgb]{0.95,0.95,0.95} 0.17\% & \cellcolor[rgb]{0.95,0.95,0.95} 0.00\% & \cellcolor[rgb]{0.95,0.95,0.95} 0.00\% & \cellcolor[rgb]{0.95,0.95,0.95} 25.57\% & \cellcolor[rgb]{0.95,0.95,0.95} 0.67\% & \cellcolor[rgb]{0.95,0.95,0.95} 0.00\% & \cellcolor[rgb]{0.95,0.95,0.95} 52.09\% & \cellcolor[rgb]{0.95,0.95,0.95} 0.00\% & \cellcolor[rgb]{0.95,0.95,0.95} 0.00\% & \cellcolor[rgb]{0.95,0.95,0.95} 0.00\%\\[-0.5pt]
\cline{2-14}\\[-9.8pt]
& Error rate & 1.000 & 1.007 & 1.023 & 1.000 & 1.000 & 1.240 & 1.000 & 1.000 & 0.799 & 1.020 & 1.000 & 1.000\\[-0.5pt]
\cline{1-14}\\[-10.0pt]
\multirow{2.2}{*}{Inverse freq. loss} & \cellcolor[rgb]{0.95,0.95,0.95} Fscore & \cellcolor[rgb]{0.95,0.95,0.95} 0.00\% & \cellcolor[rgb]{0.95,0.95,0.95} 0.02\% & \cellcolor[rgb]{0.95,0.95,0.95} 0.00\% & \cellcolor[rgb]{0.95,0.95,0.95} 0.00\% & \cellcolor[rgb]{0.95,0.95,0.95} 0.00\% & \cellcolor[rgb]{0.95,0.95,0.95} 0.00\% & \cellcolor[rgb]{0.95,0.95,0.95} 2.39\% & \cellcolor[rgb]{0.95,0.95,0.95} 0.00\% & \cellcolor[rgb]{0.95,0.95,0.95} 26.89\% & \cellcolor[rgb]{0.95,0.95,0.95} 0.00\% & \cellcolor[rgb]{0.95,0.95,0.95} 0.00\% & \cellcolor[rgb]{0.95,0.95,0.95} 0.00\% \\[-0.5pt]
\cline{2-14}
&&&&&&&&&&&&&\\[-10pt]
& Error rate & 1.000 & 1.000 & 1.000 & 1.000 & 1.000 &1.000 & 0.989 & 1.000 & 1.010 & 1.000 & 1.000 & 1.000\\[-0.5pt]
\cline{1-14}\\[-10.0pt]
\multirow{1.1}{*}{Duration robust loss} & \cellcolor[rgb]{0.95,0.95,0.95} Fscore & \cellcolor[rgb]{0.95,0.95,0.95} 0.00\% & \cellcolor[rgb]{0.95,0.95,0.95} 2.86\% & \cellcolor[rgb]{0.95,0.95,0.95} 1.20\% & \cellcolor[rgb]{0.95,0.95,0.95} 0.00\% & \cellcolor[rgb]{0.95,0.95,0.95} 0.00\% & \cellcolor[rgb]{0.95,0.95,0.95} 25.65\% & \cellcolor[rgb]{0.95,0.95,0.95} 4.01\% & \cellcolor[rgb]{0.95,0.95,0.95} 0.00\% & \cellcolor[rgb]{0.95,0.95,0.95} \textbf{55.22\%} & \cellcolor[rgb]{0.95,0.95,0.95} 0.00\% & \cellcolor[rgb]{0.95,0.95,0.95} 0.00\% & \cellcolor[rgb]{0.95,0.95,0.95} 0.00\% \\[-0.5pt]
\cline{2-14}
\multirow{0.9}{*}{($\gamma = 1.0$)}&&&&&&&&&&&&&\\[-10pt]
& Error rate & 1.000 & 1.021 & 1.045 & 1.000 & 1.000 & 1.256 & 0.988 & 1.000 & \textbf{0.768} & 1.031 & 1.000 & 1.000\\[-0.5pt]
\cline{1-14}\\[-10.0pt]
\multirow{1.1}{*}{Duration robust loss} & \cellcolor[rgb]{0.95,0.95,0.95} Fscore & \cellcolor[rgb]{0.95,0.95,0.95} 0.00\% & \cellcolor[rgb]{0.95,0.95,0.95} \textbf{7.21\%} & \cellcolor[rgb]{0.95,0.95,0.95} \textbf{2.37\%} & \cellcolor[rgb]{0.95,0.95,0.95} 0.00\% & \cellcolor[rgb]{0.95,0.95,0.95} 0.00\% & \cellcolor[rgb]{0.95,0.95,0.95} \textbf{28.41}\% & \cellcolor[rgb]{0.95,0.95,0.95} \textbf{13.35\%} & \cellcolor[rgb]{0.95,0.95,0.95} 0.00\% & \cellcolor[rgb]{0.95,0.95,0.95} 54.01\% & \cellcolor[rgb]{0.95,0.95,0.95} 0.00\% & \cellcolor[rgb]{0.95,0.95,0.95} 0.00\% & \cellcolor[rgb]{0.95,0.95,0.95} 0.00\% \\[-0.5pt]
\cline{2-14}
\multirow{0.9}{*}{($\gamma = 4.0$)}&&&&&&&&&&&&&\\[-10pt]
& Error rate & 1.000 & 1.084 & 1.092 & 1.003 & 1.000 & 1.224 & \textbf{0.960} & 1.000 & 0.811 & 1.032 & 1.000 & 1.000\\[-1pt]
\wcline{1-14}
\vspace{-2pt}
\end{tabular}
}
%
%
%\vspace{0pt}
\small
\centering
\scalebox{0.7656}[0.7656]{
\begin{tabular}{llccccccccccccc}
\wcline{1-15}
&&&&&&&&&&&&&\\[-9pt]
\multicolumn{2}{c}{\multirow{2}{*}{Event}} &\! \multirow{2}{*}{dishes} \!\!&\!\! \multirow{2}{*}{drawer} \!\!&\!\! \multirow{2}{*}{fan} \!\!&\!\! glass \!\!&\!\! keyboard \!\!&\!\! large \!\!&\!\! mouse \!\!&\!\! mouse \!\!&\!\! people \!\!&\!\! people \!\!&\!\! washing \!\!&\!\! water tap \!\!&\!\! wind\!\!\\[-1pt]
&&\!\!  \!\!&\!\!  \!\!&\!\!  \!\!&\!\! jingling \!\!&\!\! typing \!\!&\!\! vehicle \!\!&\!\! clicking \!\!&\!\! wheeling \!\!&\!\! talking \!\!&\!\! walking \!\!&\!\! dishes \!\!&\!\! running \!\!&\!\! blowing\!\\[-0.5pt]
\wcline{1-15}\\[-9.5pt]
\multirow{2.2}{*}{BCE loss} & \cellcolor[rgb]{0.95,0.95,0.95} Fscore &\! \cellcolor[rgb]{0.95,0.95,0.95} 0.00\% \!\!&\!\! \cellcolor[rgb]{0.95,0.95,0.95} 0.00\% \!\!&\!\! \cellcolor[rgb]{0.95,0.95,0.95} 17.86\% \!\!&\!\! \cellcolor[rgb]{0.95,0.95,0.95} 0.00\% \!\!&\!\! \cellcolor[rgb]{0.95,0.95,0.95} 0.01\% \!\!&\!\! \cellcolor[rgb]{0.95,0.95,0.95} 45.25\% \!\!&\!\! \cellcolor[rgb]{0.95,0.95,0.95} 0.00\% \!\!&\!\! \cellcolor[rgb]{0.95,0.95,0.95} 0.00\% \!\!&\!\! \cellcolor[rgb]{0.95,0.95,0.95} 2.68\% \!\!&\!\! \cellcolor[rgb]{0.95,0.95,0.95} 26.32\% \!\!&\!\! \cellcolor[rgb]{0.95,0.95,0.95} 15.63\% \!\!&\!\! \cellcolor[rgb]{0.95,0.95,0.95} 13.00\% \!\!&\!\! \cellcolor[rgb]{0.95,0.95,0.95} 0.00\% \!\\[-0.5pt]
\cline{2-15}\\[-10pt]
& Error rate &\! 1.000 \!\!&\!\! 1.000 \!\!&\!\! 0.908 \!\!&\!\! 1.000 \!\!&\!\! 1.000 \!\!&\!\! 1.202 \!\!&\!\! 1.000 \!\!&\!\! 1.000 \!\!&\!\! 1.053 \!\!&\!\! 0.907 \!\!&\!\! \textbf{0.959} \!\!&\!\! 0.942 \!\!&\!\! 1.000 \!\\[-0.5pt]
\cline{1-15}\\[-10.0pt]
\multirow{2.2}{*}{Inverse freq. loss} & \cellcolor[rgb]{0.95,0.95,0.95} Fscore &\! \cellcolor[rgb]{0.95,0.95,0.95} 0.00\% \!\!&\!\! \cellcolor[rgb]{0.95,0.95,0.95} 0.00\% \!\!&\!\! \cellcolor[rgb]{0.95,0.95,0.95} 0.00\% \!\!&\!\! \cellcolor[rgb]{0.95,0.95,0.95} 0.00\% \!\!&\!\! \cellcolor[rgb]{0.95,0.95,0.95} 0.00\% \!\!&\!\! \cellcolor[rgb]{0.95,0.95,0.95} 10.93\% \!\!&\!\! \cellcolor[rgb]{0.95,0.95,0.95} 0.00\% \!\!&\!\! \cellcolor[rgb]{0.95,0.95,0.95} 0.00\% \!\!&\!\! \cellcolor[rgb]{0.95,0.95,0.95} 0.00\% \!\!&\!\! \cellcolor[rgb]{0.95,0.95,0.95} 0.00\% \!\!&\!\! \cellcolor[rgb]{0.95,0.95,0.95} 0.00\% \!\!&\!\! \cellcolor[rgb]{0.95,0.95,0.95} 2.73\% \!\!&\!\! \cellcolor[rgb]{0.95,0.95,0.95} 0.00\% \!\\[-0.5pt]
\cline{2-15}
&&&&&&&&&&&&&\!\\[-10pt]
& Error rate &\! 1.000 \!\!&\!\! 1.000 \!\!&\!\! 1.000 \!\!&\!\! 1.000 \!\!&\!\! 1.000 \!\!&\!\! \textbf{0.968} \!\!&\!\! 1.000 \!\!&\!\! 1.000 \!\!&\!\! 1.000 \!\!&\!\! 1.000 \!\!&\!\! 1.000 \!\!&\!\! 0.986 \!\!&\!\! 1.000 \!\\[-0.5pt]
\cline{1-15}\\[-10.0pt]
\multirow{1.1}{*}{Duration robust loss} & \cellcolor[rgb]{0.95,0.95,0.95} Fscore &\! \cellcolor[rgb]{0.95,0.95,0.95} 0.01\% \!\!&\!\! \cellcolor[rgb]{0.95,0.95,0.95} 0.00\% \!\!&\!\! \cellcolor[rgb]{0.95,0.95,0.95} 25.44\% \!\!&\!\! \cellcolor[rgb]{0.95,0.95,0.95} 0.00\% \!\!&\!\! \cellcolor[rgb]{0.95,0.95,0.95} 0.23\% \!\!&\!\! \cellcolor[rgb]{0.95,0.95,0.95} \textbf{47.67\%} \!\!&\!\! \cellcolor[rgb]{0.95,0.95,0.95} 0.00\% \!\!&\!\! \cellcolor[rgb]{0.95,0.95,0.95} 0.00\% \!\!&\!\! \cellcolor[rgb]{0.95,0.95,0.95} 4.40\% \!\!&\!\! \cellcolor[rgb]{0.95,0.95,0.95} \textbf{32.09\%} \!\!&\!\! \cellcolor[rgb]{0.95,0.95,0.95} 13.97\% \!\!&\!\! \cellcolor[rgb]{0.95,0.95,0.95} 34.16\% \!\!&\!\! \cellcolor[rgb]{0.95,0.95,0.95} 0.00\% \!\\[-0.5pt]
\cline{2-15}
\multirow{0.1}{*}{($\gamma = 1.0$)}&&&&&&&&&&&&&\\[-10pt]
& Error rate &\! 1.000 \!\!&\!\! 1.000 \!\!&\!\! 0.861 \!\!&\!\! 1.000 \!\!&\!\! 1.000 \!\!&\!\! 1.190 \!\!&\!\! 1.000 \!\!&\!\! 1.000 \!\!&\!\! 1.091 \!\!&\!\! \textbf{0.898} \!\!&\!\! 1.003 \!\!&\!\! 0.801 \!\!&\!\! 1.000 \!\\[-0.5pt]
\cline{1-15}\\[-10.0pt]
\multirow{1.1}{*}{Duration robust loss} & \cellcolor[rgb]{0.95,0.95,0.95} Fscore &\! \cellcolor[rgb]{0.95,0.95,0.95} \textbf{1.66\%} \!\!&\!\! \cellcolor[rgb]{0.95,0.95,0.95} \textbf{0.02\%} \!\!&\!\! \cellcolor[rgb]{0.95,0.95,0.95} \textbf{28.91\%} \!\!&\!\! \cellcolor[rgb]{0.95,0.95,0.95} 0.00\% \!\!&\!\! \cellcolor[rgb]{0.95,0.95,0.95} \textbf{0.49\%} \!\!&\!\! \cellcolor[rgb]{0.95,0.95,0.95} 47.44\% \!\!&\!\! \cellcolor[rgb]{0.95,0.95,0.95} \textbf{0.08\%} \!\!&\!\! \cellcolor[rgb]{0.95,0.95,0.95} 0.00\% \!\!&\!\! \cellcolor[rgb]{0.95,0.95,0.95} \textbf{5.26\%} \!\!&\!\! \cellcolor[rgb]{0.95,0.95,0.95} 30.34\% \!\!&\!\! \cellcolor[rgb]{0.95,0.95,0.95} \textbf{20.65\%} \!\!&\!\! \cellcolor[rgb]{0.95,0.95,0.95} \textbf{38.76\%} \!\!&\!\! \cellcolor[rgb]{0.95,0.95,0.95} \textbf{0.02\%} \!\\[-0.5pt]
\cline{2-15}
\multirow{0.1}{*}{($\gamma = 4.0$)}&&&&&&&&&&&&&\\[-10pt]
& Error rate &\! \textbf{0.998} \!\!&\!\! 1.000 \!\!&\!\! \textbf{0.839} \!\!&\!\! 1.000 \!\!&\!\! 1.000 \!\!&\!\! 1.141 \!\!&\!\! 1.000 \!\!&\!\! 1.000 \!\!&\!\! 1.119 \!\!&\!\! 0.907 \!\!&\!\! 1.031 \!\!&\!\! \textbf{0.773} \!\!&\!\! 1.024 \!\\[-1pt]
\wcline{1-15}
\end{tabular}
}
\vspace{10pt}
\end{table*}

One simple idea to address the imbalanced-data problem is classwise reweighting of the loss function in accordance with the inverse frequency of sound event occurrences as follows.

\vspace{-10pt}
\begin{align}
{\rm E}_{2} ({\bi \theta}) &= - \sum^{M}_{m=1} \frac{C}{N_{m} + C} \sum^{N}_{n=1} \! {\Big \{} z_{n,m} \log {\big (} s(y_{n,m}) {\big )} \nonumber\\[0pt]
&\hspace{25pt} + (1 - z_{n,m}) \log {\big (} 1 - s(y_{n,m}) {\big )} {\Big \}}
\label{eq:invfreq_loss}
\vspace{-10pt}
\end{align}
%\vspace{0pt}
%

\noindent Here, $N_{m}$ and $C$ are the number of frames of sound event $m$ in a sound clip and a constant number, respectively.
However, in a preliminary experiment, we confirmed that the simple classwise reweighting using ${\rm E}_{2}({\bi \theta})$ is not effective for SED.
%
%---------------------------------------------------
%\vspace{0pt}
\section{Proposed Method}
\label{Proposed}
%\vspace{0pt}
%---------------------------------------------------
In this paper, we propose a duration robust loss function for SED, whereby training can be focused on sound events of short duration.
In the proposed method, we focus on the relationship between the sound event duration and the ease/difficulty of model training.
In particular, many sound events of long duration (e.g., ``fan'' and ``car'') are stationary sounds, which have less variation in their acoustic features and their model training is easy.
Meanwhile, some sound events of short duration (e.g., ``object impact'' and ``keyboard typing'') have more than one audio pattern, such as attack, decay, and release parts.
Therefore, the ease/difficulty of model training is important information for controlling the training weight, as well as a more direct way of controlling the loss contribution.

To control the training weight of sound events in accordance with the ease/difficulty of model training, we add factors $(1 - s(y_{n,m}))^{\gamma}$ and $s(y_{n,m})^{\gamma}$ to the BCE loss as follows:

\vspace{-8pt}
\begin{align}
{\rm E}_{3} ({\bi \theta}) &= - \sum^{M}_{m=1} \sum^{N}_{n=1} \! {\Big \{} (1 - s(y_{n,m}))^{\gamma} \hspace{1pt} z_{n,m} \log {\big (} s(y_{n,m}) {\big )} \nonumber\\[0pt]
&\hspace{10pt} + s(y_{n,m})^{\gamma} \hspace{1pt} (1 - z_{n,m}) \log {\big (} 1 - s(y_{n,m}) {\big )} {\Big \}},
\label{eq:prop_loss}
%\vspace{0pt}
\end{align}
%\vspace{0pt}
%

\noindent where $\gamma$ is the weighting parameter that controls the focusing weight.
When sound event $m$ is active in time frame $n$ but the network output $s(y_{n,m})$ is a small value, the reweighting factor $(1 - s(y_{n,m}))^{\gamma}$ does not greatly affect the loss, whereas if the network output is a large value, the reweighting factor approaches zero and the loss is down-weighted.
Thus, the loss function focuses the model training on the sound event classes that are difficult to train.

It is considered that the concept of the duration robust loss function is similar to that of the focal loss \cite{Lin_ICCV2017_01} in objective detection.
In objective detection, there is also a serious imbalance problem, where background samples tend to have a large number of pixels with similar patterns, whereas the foreground samples are likely to have a relatively small number of pixels.
\begin{table}[t]
\vspace{-15pt}
\small
\caption{Experimental conditions}
\vspace{-17pt}
\label{tbl:parameter}
\begin{center}
\begin{tabular}{ll}
\wcline{1-2}
&\\[-9pt]
\!Acoustic feature \!\!&\!\! Log mel-band energy (64 dim.)\!\!\\
\!Frame length \hspace{-3pt} / \hspace{-3pt} shift \!\!&\!\! 40 ms \hspace{-3pt} / \hspace{-3pt} 20 ms\!\!\\
\!Length of sound clip \!\!&\!\! 10 s\!\!\\
\cline{1-2}
\!&\\[-9pt]
\!Network structure \!\!&\!\! 3 CNN $+$ 1 BiGRU $+$ 1 fully conn.\!\!\\[1pt]
\!\# channels of CNN layers \!\!&\!\! 128, 128, 128\!\!\\[0pt]
\!Filter size \!\!&\!\! 3$\times$3, 3$\times$3, 3$\times$3\!\!\\[0pt]
\!Pooling size \!\!&\!\! 1$\times$8, 1$\times$4, 1$\times$2 (max pooling)\!\!\\[0pt]
\!\# units in GRU layer \!\!&\!\! 32\!\!\\[0pt]
\!\# units in fully conn. layer \!\!&\!\! 32\!\!\\[0pt]
\!Detection threshold\!\!&\!\! 0.5\!\!\\[0pt]
\!Constant number $C$\!\!&\!\! 500\!\!\\
\wcline{1-2}
\end{tabular}
%\vspace{0pt}
\end{center}
\end{table}
\begin{table}[t]
\vspace{-13pt}
\caption{Average performance of SED}
\vspace{2pt}
\small
\centering
\begin{tabular}{lcc}
\wcline{1-3}
&&\\[-9pt]
\multicolumn{1}{c}{Method} \!\!\!\!&\!\! Macro-Fscore &\!\! Micro-Fscore\\[-1pt]
\wcline{1-3}
&&\\[-9pt]
BCE loss \!\!\!\!\!\!& \multicolumn{1}{r}{8.01\%} \ \ \ \ & \multicolumn{1}{r}{\!\! 27.83\%} \ \ \ \ \\[0pt]
Inverse frequency loss \!\!\!\!\!\!& \multicolumn{1}{r}{1.72\%} \ \ \ \ & \multicolumn{1}{r}{\!\! 7.44\%} \ \ \ \ \\[0pt]
Duration robust loss ($\gamma=1.0$) \!\!\!\!\!\!& \multicolumn{1}{r}{9.88\%} \ \ \ \ & \multicolumn{1}{r}{\!\! 31.40\%} \ \ \ \ \\[0pt]
Duration robust loss ($\gamma=4.0$) \!\!\!\!\!& \multicolumn{1}{r}{\textbf{11.16\%}} \ \ \ \ & \multicolumn{1}{r}{\!\! \textbf{32.20\%}} \ \ \ \ \\[-1pt]
\wcline{1-3}
\end{tabular}
\vspace{15pt}
\label{tbl:performance01}
\end{table}
%
%
%
%---------------------------------------------------
%\vspace{0pt}
\section{Experiments}
\label{sec:experiments}
%\vspace{0pt}
%---------------------------------------------------
%- - - - - - - - - - - - - - - - - - - - - - - - - - -
%\vspace{0pt}
\subsection{Experimental Conditions}
\label{ssec:conditions}
%\vspace{0pt}
%- - - - - - - - - - - - - - - - - - - - - - - - - - -
We evaluated the performance of SED using the proposed duration robust loss function.
As an evaluation dataset, we constructed the dataset composed of parts of TUT Sound Events 2016 development, 2017 development, and TUT Acoustic Scenes 2016 development \cite{Mesaros_EUSIPCO2016_01,Mesaros_DCASE2017_01}.
We selected a total of 192 min of sound clips including the 25 types of sound event listed in Fig.~\ref{fig:num_frame_01}.
As shown in Fig.~\ref{fig:num_frame_01}, the number of time frames between sound events is seriously unbalanced, e.g., sound event ``fan'' makes up a total of more than 100,000 frames in the dataset, while sound event ``mouse clicking'' accounts for less than 3,500 frames.
Note that the datasets were recorded not for the detection task of rare sound but for the analysis of real-life sounds; thus, the analysis of seriously imbalanced data is a general problem in SED.

As an acoustic feature, we used the 64-dimensional log mel-band energy, which was calculated every 40 ms with a 20 ms frame shift.
The acoustic feature was fed to the neural network with 3 CNN layers, 1 bidirectional gated recurrent unit (GRU) layer, and 1 fully connected layer, which was used for the baseline system of the DCASE2018 challenge task 4 \cite{Serizel_DCASE2018_01}.
The performance of sound event detection was evaluated using the segment-based macro- and micro-Fscores \cite{Mesaros_AS2016_01}.
Other experimental conditions are listed in Table~\ref{tbl:parameter}.
%
%
%- - - - - - - - - - - - - - - - - - - - - - - - - - -
%\vspace{0pt}
\subsection{Experimental Results}
\label{ssec:results}
%\vspace{0pt}
%- - - - - - - - - - - - - - - - - - - - - - - - - - -
Table~\ref{tbl:performance01} shows the average performances of SED using the BCE loss, the BCE loss with the inverse frequency class reweighting (referred to as inverse frequency loss), and the proposed duration robust loss.
For each loss, we conducted the evaluation experiment 10 times with random initial values for model parameters.
The results show that the proposed SED method improves both the macro- and micro-Fscores by 3.15 and 4.37 percentage points, respectively, compared with SED using the BCE loss function.
Because the macro-Fscore tends to be weighted towards a sound event class that has a small number of frames, the results indicate that the proposed duration loss function is effective for the sound events of short length.
On the other hand, the micro-Fscore is likely to be weighted towards a sound event class with a large number of frames; thus, the experimental results also indicate that the proposed method can balance the detection of sound events of both short and long durations.

\begin{figure}[t]
%\vspace{0pt}
\centering
%\vspace{0pt}
\includegraphics[width=0.99\columnwidth]{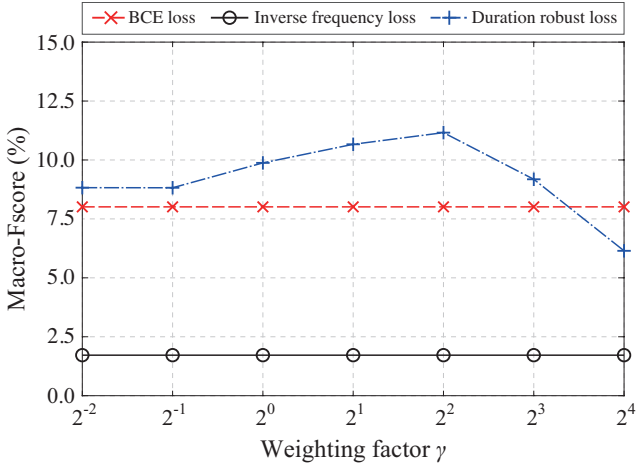}
\vspace{-4pt}
\caption{Average macro-Fscores of SED with various weighting factors $\gamma$}
\label{fig:macro_fscore}
\vspace{4pt}
\end{figure}

To investigate the details of SED performance, we show the detection results for each sound event in Table~\ref{tbl:each_event}.
The results show that the proposed method improves both the Fscore and error rate for many sound events.
For instance, the sound events ``(object) impact,'' ``(object) rustling,'' ``dishes,'' and ``keyboard typing,'' which have short durations (0.2--1.5 s), can be detected more precisely by the proposed method.
Similarly, the detection performance of sound events ``bird singing,'' ``car,'' ``can,'' and ``large vehicle,'' which have long durations, is also improved.
On the other hand, several sound events with very short durations (e.g., mouse wheeling) cannot be detected even by the proposed method; thus, this must be addressed in the future.

Figs.~\ref{fig:macro_fscore} and \ref{fig:micro_fscore} shows the average macro- and micro-Fscores, respectively, for various weighting factors $\gamma$.
The results show that even when the weighting factor $\gamma$ is changed from $2^{-2}$ to $2^{2}$, the proposed method achieves better results than conventional methods in terms of both macro- and micro-Fscores.
Thus, the experimental results show that incorporating duration robust loss leads to stable performance with various weighting factors $\gamma$.
%
%
%
%---------------------------------------------------
%\vspace{0pt}
\section{Conclusion}
\label{sec:conclusion}
%\vspace{0pt}
%---------------------------------------------------
In this paper, we proposed sound event detection using the duration robust loss function, which can focus training on sound events of short duration.
In the proposed method, we assumed that many sound events of long duration are stationary sounds, which have less variation in their acoustic features and their model training is easy.
On the basis of the ease/difficulty of model training, we apply the reweighting factor to the BCE loss, which focuses the model training on the sound event classes for which model training is difficult.
Experimental results obtained using the imbalanced dataset in sound event classes indicate that the proposed method can detect sound events of short durations more precisely.
%
%
%-----------------------
%\vspace{0pt}
\section{Acknowledgement}
\label{sec:ack}
%\vspace{0pt}
%-----------------------
This work was supported by JSPS KAKENHI Grant Number JP19K20304 and NVIDIA GPU Grant Program.
\begin{figure}[t!]
%\vspace{0pt}
\centering
\includegraphics[width=0.99\columnwidth]{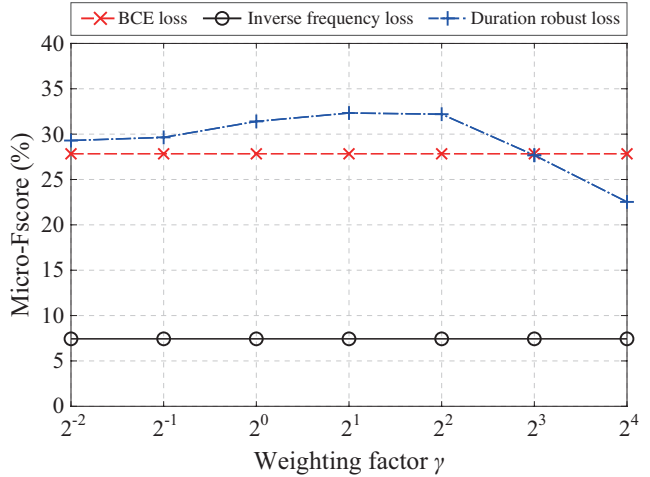}
\vspace{-4pt}
\caption{Average micro-Fscores of SED with various weighting factors $\gamma$}
\label{fig:micro_fscore}
\vspace{4pt}
\end{figure}
%
%
%
% -------------------------------------------------------------------------
% Either list references using the bibliography style file IEEEtran.bst
\bibliographystyle{IEEEtran}
\bibliography{IEEEabrv,DCASE2020refs,KeisukeImoto10}
% -------------------------------------------------------------------------
%
\end{sloppy}
\end{document}